\documentclass[conference, 10pt]{IEEEtran}

\usepackage{cite}
\usepackage{url}
\usepackage{xcolor}
\usepackage{graphicx}
\usepackage{color}
\usepackage{placeins}
\usepackage{float}
\usepackage{tabularx,colortbl}
\usepackage{ifthen}
\usepackage{amsmath}
\usepackage{amsfonts}
\usepackage{pgfplots}
\usepackage{algorithmic}
\usepackage{lipsum}
\usepackage{algorithm}
\usepackage{color}
\usepackage{array}
\usepackage[font=footnotesize]{caption}
\usepackage{textcomp}
\usepackage{stfloats}
\usepackage{url}
\usepackage{verbatim}
\usepackage{graphicx}
\usepackage{adjustbox}
\usepackage{cite}
\usepackage{soul}
\usepackage{multicol}
\usepackage{subcaption}
\usepackage{multirow}
\usepackage{booktabs}
\usepackage{titlesec}

\usepackage{tikz}
\usepackage{pgfplots}
\usepackage{subcaption}
\usepackage{threeparttablex}
\usepackage[all]{nowidow}
\usepackage{lipsum} 
\def\BibTeX{{\rm B\kern-.05em{\sc i\kern-.025em b}\kern-.08em
    T\kern-.1667em\lower.7ex\hbox{E}\kern-.125emX}}

\makeatletter

\newcounter{author}
\renewcommand{\author}[2][]{
   \stepcounter{author}
   \@namedef{author@\theauthor}{#2}
   \@namedef{authorlabel@\theauthor}{#1}
}

\newcounter{address}
\newcommand{\address}[2][]{
   \stepcounter{address}
   \@namedef{address@\theaddress}{#2}
   \@namedef{addresslabel@\theaddress}{#1}
}

\newcommand{\alsep}{and}

\def\newmaketitle{\par%
  \begingroup%
  \normalfont%
  \def\thefootnote{}
  \def\footnotemark{}
  \let\@makefnmark\relax
  \footnotesize
  \footnotesep 0.7\baselineskip
  \normalsize%
  \twocolumn[\thenewmaketitle\@IEEEaftertitletext]%
  \if@IEEEusingpubid
     \enlargethispage{-\@IEEEpubidpullup}%
  \fi
  \endgroup
  \setcounter{footnote}{0}\let\maketitle\relax\let\@maketitle\relax
  \gdef\@thanks{}%
  \let\thanks\relax}

\def\thenewmaketitle{
  \newpage
  \begin{center}%
    \vskip0.2em{\Huge\@IEEEcompsoconly{\sffamily}\@IEEEcompsocconfonly{\normalfont\normalsize\vskip 2\@IEEEnormalsizeunitybaselineskip
   \bfseries\large}\@title\par}\vskip1.0em\par%
    \vspace{1ex}
    \newcounter{c@author}
    \newcounter{c@tmp}
    \ifthenelse{\value{author}=2}{%
      \newcommand{\liand}{ and }}{%
      \newcommand{\liand}{, and }}
    \ifthenelse{\value{address}<2}{%
      \@nameuse{author@1}%
      \stepcounter{c@author}%
      \whiledo{\value{c@author}<\value{author}}{%
        \setcounter{c@tmp}{\value{author}}%
        \addtocounter{c@tmp}{-\value{c@author}}%
        \ifthenelse{\value{c@tmp}=1}{%
          \renewcommand{\alsep}{\liand}}{\renewcommand{\alsep}{, }}%
        \stepcounter{c@author}\alsep \@nameuse{author@\thec@author}}\\%
    }
    {
      \@nameuse{author@1}${}^{(\ref{\@nameuse{authorlabel@1}})}$%
      \stepcounter{c@author}%
      \whiledo{\value{c@author}<\value{author}}{%
      \setcounter{c@tmp}{\value{author}}%
      \addtocounter{c@tmp}{-\value{c@author}}%
      \ifthenelse{\value{c@tmp}=1}{%
        \renewcommand{\alsep}{\liand}}{\renewcommand{\alsep}{, }}%
      \stepcounter{c@author}\alsep \@nameuse{author@\thec@author}%
        ${}^{(\ref{\@nameuse{authorlabel@\thec@author}})}$%
      }
    }
    \vspace{0.2ex}

    \ifthenelse{\value{address}>0}{%
      \ifthenelse{\value{address}=1}{
        {\@nameuse{address@1}}
      }
      {
        \newcounter{c@address}

        \begin{center}
        \whiledo{\value{c@address}<\value{address}}
        {
          \refstepcounter{c@address}
            ${}^{(\thec@address)}$\,%
              \label{\@nameuse{addresslabel@\thec@address}}%
              \@nameuse{address@\thec@address}\\ %
        }
        \end{center}
      } 
    }
    {
      \relax
    }
  \end{center}
}

\makeatother

\title{Feasibility Study of Curvature Effect in Flexible Antenna Arrays for 2-Dimensional Beam Alignment of 6G Wireless Systems}

\author[org1]{Mahdi Alesheikh}
\author[org2]{Soheil Saadat}
\author[org1]{Hamidreza Aghasi}
\address[org1]{The University of California, Irvine, CA, 92617 USA}
\address[org2]{MFLEX Inc., Irvine, CA,  92617, USA}

\begin{document}

\newmaketitle

\begin{abstract}
This article investigates the influential role of flexible antenna array curvature on the performance of 6G communication systems with carrier frequencies above 100 GHz. It is demonstrated that the curvature of flexible antenna arrays can be leveraged for 2-dimensional beam alignment in phased arrays with relatively small insertion loss. The effect of antenna array bending on the radiation properties such as gain and antenna impedance are analytically studied and simulated for a 4$\times$4 microstrip patch antenna array operating between 97.5-102.5 GHz. Moreover, the deployment of this flexible antenna array is examined for 6G wireless transceivers based on 65nm CMOS technology and simulated for three variants of quadrature amplitude modulation (4QAM, 16 QAM, and 64 QAM). The communication performance in terms of signal-to-noise ratio (SNR) and bit error rate (BER) is evaluated using analytical derivations and simulation results which exhibit a close match.  
\end{abstract}

\vspace{-0.1in}
\section{Introduction}
\IEEEPARstart{I}{n} light of advances in semiconductor technologies, the demand for higher data rates in communication systems and higher resolution in imaging and radar technologies is continuously growing
\cite{liu}. By increasing the carrier frequency and signal bandwidth, the communication data rate and sensing resolution can increase, respectively 
\cite{Journal_TCASI_RFQAMTX_MOveisi}.
To extend the angular coverage of mm-wave systems, phased-array architectures are adopted. For mm-wave phased array systems, one major challenge is the beam alignment between transmitter (TX) and receiver (RX) channels, which is currently done by various phase shifting techniques\cite{beam_allign}. A common challenge among all the techniques is the design of low-power broadband phase-shifting/time-delay circuits with relatively small insertion loss \cite{phase_shifter_Wooram}. Not only does the dynamic range of these phase shifters impact the beam alignment range, but they also consume considerable power when realized by active devices\cite{18}.
\setlength{\belowcaptionskip}{-5pt} 
 \begin{figure}
       \centering
       \includegraphics[width=0.85\linewidth]{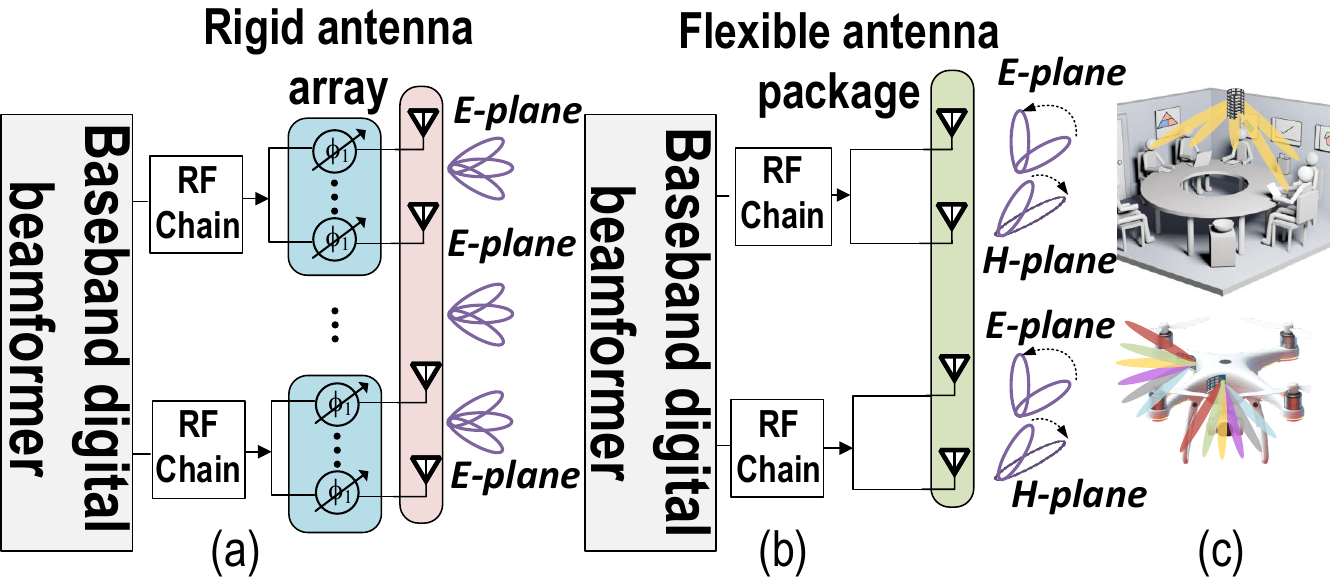}
       \caption{(a) conventional 1-D beam alignment hybrid beamforming architecture, (b) proposed 2-D beam alignment architecture by extended aerial coverage, (c) target applications.}
       \label{fig:1}
   \end{figure} 

By adopting the mm-wave phased arrays for multiple input multiple output communication systems, phase shifters in each TX and RX channel should be tuned in order to achieve beam alignment and extended coverage \cite{hybrid}. However, as shown in Fig. 1(a) the beam alignment of these conventional linear array systems is limited by the phase shifters. Building on top of recent advances in fabrication of mm-wave antenna elements above 100 GHz using flexible printed circuit technologies \cite{Hedayat}, this paper delves into the analysis, simulation, and investigation of extended 2-D beam alignment by deploying the curvature of flexible antenna arrays as shown in Fig. \ref{fig:1}(b) for indoor and outdoor applications, Fig. 1(c). The main objective of the proposed design is to get rid of the unwanted insertion loss of phase shifters in mm-wave phased arrays and achieve beam alignment with substantially reduced insertion loss. Using the same technology that was deployed in \cite{Hedayat}, the beam rotation in a 4$\times$4 planar microstrip patch antenna array at 97.5-102.5 GHz is analytically studied in Section II where the relationship between the antenna parameters and the degree of folding is presented. In Section III the presented antenna array is deployed in a 6G wireless transceiver based on a 65-nm CMOS transciever and simulations are conducted and compared with analytical derivations for the performance of the 6G systems. The paper is concluded in Section IV.

\section{Flexible Antenna Design}
To design a multiple-input multiple-output beamforming phased-array, the packaging of the antenna, system-on-chip (SOC), the main board and interposer should be considered. We propose two multi-level integration schemes that are custom designed for this specific application and are shown in Fig. \ref{fig:package}. In Fig. \ref{fig:package}(a) each CMOS transceiver chip and the corresponding interposer will be placed between two flexible layers, modified polyimide material (MPI), with $\epsilon_r=3.1$ and loss tangent of 0.003) at the bottom and top. This scheme is preferred for MIMO systems. In Fig. \ref{fig:package}(b), the buried chip inside the flexible printed circuit board (FPC) materials is proposed with reduced insertion loss for the interface between the board and the chip. Both technologies have been experimentally validated at MFLEX Inc. and according to our recent measurements in Fig. \ref{fig:package}(c,d), for various thicknesses of the substrate (MPI), the minimum bend radius of few mm is achievable. Moreover, the maximum bend angle of the board increases by the dimensions of the design and can extend to 330 degrees for a board with 25 mm dimension and 250 $\mu$m thickness. By taking into account these technology constraints, the analytical derivations and simulation results are provided to capture the effect of 2-D folding in 6G antenna arrays.


\subsection{3-D Folding of a Patch Antenna Array}
 \begin{figure}
       \centering
       \includegraphics[trim={0.3cm 0.0cm 0.0cm 0.0cm},clip, width=0.9\columnwidth]{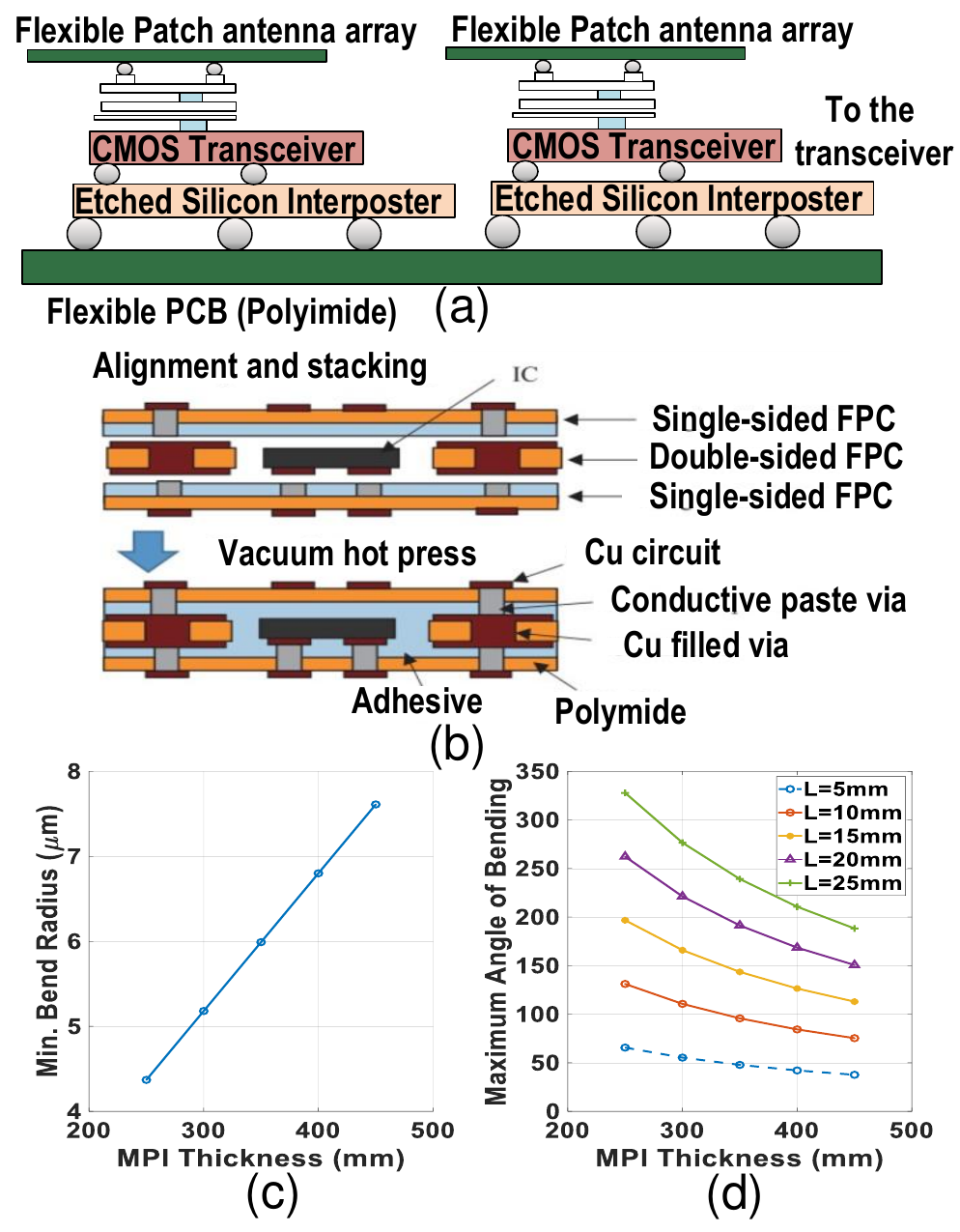}
       \caption{(a) heterogeneous package comprising CMOS chipsets and antenna on FPC, (b) buried chip in FPC package. Measured values of (c) bend radius for various substrate thickness, and (d) board flexibility maximum angle vs. its dimensions.}
       \label{fig:package}
   \end{figure}

\begin{figure}
    \centering
    \includegraphics[width=0.9\linewidth]{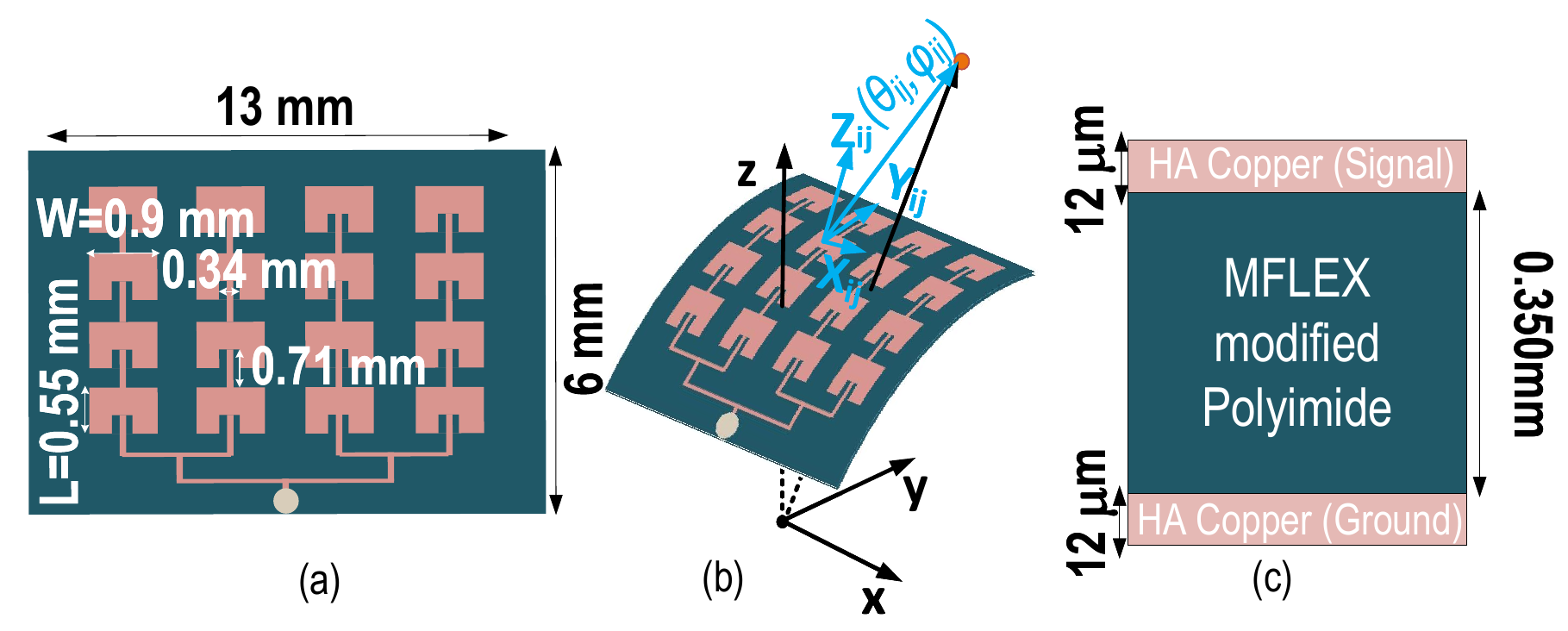}
    \caption{(a) Unfolded rectangular patch antenna array, (b) illustration of array bending, (c) metal layer details of the flexible patch antenna array.}
    \label{REF}
\end{figure}

For a patch antenna array with \textit{m} rows and \textit{n} columns, depicted in Fig. \ref{REF}(a) (\textit{m}=\textit{n}=4), the electric field associated with the arbitrary antenna element located in the \textit{i}-th row and the \textit{j}-th column is calculated with respect to its local coordinate system for an observer at ($r_{ij},\theta_{ij},\phi_{ij}$) given by \cite{Book_Antenna_Warren}:
 \vspace{-0.05in}
\begin{equation}
\small E^{l}_{\theta_{ij}} = E_0(r_{ij}) \cos\phi_{ij}  \frac{\sin\gamma_{ij}}{\gamma_{ij}}  \cos\rho_{ij}
\end{equation}
\vspace{-0.19in}
\begin{equation}
\small E^{l}_{\phi_{ij}} = -E_0(r_{ij}) \cos\theta_{ij} \sin\phi_{ij}  \frac{\sin\gamma_{ij}}{\gamma_{ij}}  \cos\rho_{ij}
\end{equation}
\vspace{-0.18in}
\begin{equation}
   \small  \gamma_{ij} = \beta  \frac{W}{2}  \sin\theta_{ij} \sin\phi_{ij} ,\rho_{ij} = \beta  \frac{L}{2}  \sin\theta_{ij} \cos\phi_{ij}
\end{equation}
\vspace{-0.2in}

which $W$, $L$,  $E_0(r)$, are the width and length and strength of electric field associated with incident power at distance $r$ from the antenna. The exponent $l$ indicates the calculation of the field vectors with respect to the local coordinate system. Moreover, $\theta_{ij}$ and $\phi_{ij}$ are the angles from patch element at $ij$-th location to the target point in the $\theta$ plane and $\phi$ plane. Variables $\gamma_{ij}$ and $\rho_{ij}$ are auxiliary for simplifying the derivation.
The magnetic fields follow a similar form with a scaling factor $\eta$ which is the intrinsic air impedance.
\setlength{\textfloatsep}{8pt} 

The 3-D folding of the antenna array will cause variations in the exact positions and orientations of the array elements. These variations will alter each radiation element's contribution to the array beam, and subsequently the radiation properties\cite{saeed}. To capture the arbitrary rotations, a change of coordinate system from spherical to cartesian, allows us to calculate basic rotation matrices $R_{x,ij}$, $R_{y,ij}$ and $R_{z,ij}$. Also, the total $E$ and $H$ vecotr fields in the Cartesian coordinate system can be written as
\begin{equation}
    \vec{E}_{ij}(\theta_{ij},\phi_{ij})=\vec{E}_{x_{ij}}(\theta_{ij},\phi_{ij})+\vec{E}_{y_{ij}}(\theta_{ij},\phi_{ij})+\vec{E}_{z_{ij}}(\theta_{ij},\phi_{ij})
\end{equation}
where $x_{ij}$, $y_{ij}$, and $z_{ij}$ are the Cartesian coordinates of the target located in the spherical coordinate of ($r_{ij}$, $\theta_{ij}$, $\phi_{ij}$).

For the $m\times n$ antenna array, the angles $\alpha_{x,ij}$, $\alpha_{y,ij}$, and $\alpha_{z,ij}$ represent the corresponding rotation of Cartesian unit vectors at the local coordinate systems when transitioning from the unfolded state to folded state.

In the next step,
$E_{ij}(\theta_{ij},\phi_{ij})$ in the global coordinate system should be calculated by applying the $R^{-1}_{tot,ij}$ on the Cartesian unit vectors of electric fields ($E^{x^l}_{ij}$,$E^{y^l}_{ij}$,$E^{z^l}_{ij}$) cascaded with the change of basis (from spherical to cartesian) matrix $T_{CB}$, i.e.,

\begin{equation}
 \vec{E}_{ij}(\theta_{ij},\phi_{ij})=R^{-1}_{tot,ij} T_{CB}[0, E_{\theta}, E_{\phi}]^T
 \label{eq:Eind}
\end{equation}

 \begin{equation}
 \small 
 \begin{aligned}
\vec{E}_{tot}(\theta_{ij},\phi_{ij})&= E_0 \sum_{i=1}^{m} \sum_{j=1}^{n} \frac{I_{ij}\sin\gamma_{ij}}{\gamma_{ij}}\cos\rho_{ij} \times \\ 
 &\quad\left[ (\mu_1)\hat{x} + (\mu_2)\hat{y} + (\mu_3)\hat{z} \right]
 \end{aligned}
 \label{fig:Efield}
\end{equation}


 Where $\vec{E}_{tot}$ is the summation of the electric fields from all individual antennas calculated based on (\ref{eq:Eind}), and $I_{ij}$ is the excitation current of \textit{ij}-th antenna and:
 \newline
 \small{
 $\mu_1=2\cos\theta_{ij} (c_x^{ij}  c_z^{ij} - s_x^{ij} s_y^{ij} s_z^{ij}),
 + \cos\phi_{ij}\sin\theta_{ij} c_y^{ij} s_z^{ij},$ 
 \newline
 $\mu_2=-2\cos\theta_{ij}s_x^{ij} c_y^{ij} - \cos\phi_{ij}\sin\theta_{ij} s_y^{ij}$,
 \newline
 $\mu_3 = 2\cos\theta_{ij} (c_x^{ij} s_z^{ij} + s_x^{ij} s_y^{ij} c_z^{ij})
  -\cos\phi_{ij}\sin\theta_{ij} c_y^{ij} c_z^{ij}.$
  }
\newline 
\normalsize
Where $s_w^{ij} = \sin \alpha_{w,ij}$ and  $c_w^{ij} = \cos \alpha_{w,ij}$ for $w\in\{x,y,z\}$. The total radiated power of the antenna array based on $\vec{E}\times\vec{H}$ is calculated by approximating $\sin \gamma_{ij}\simeq \gamma_{ij}$ and $\cos \rho_{ij}\simeq 1 $ as \cite{book_microstrip}:
 \begin{equation}
 \small
 \begin{aligned}
   P_{tot}&= \frac{E_0^2}{2 \eta} \int \sum_{i=1}^{m} \sum_{j=1}^{n} \frac{I_{ij}^2\sin\gamma_{ij}^2}{\gamma_{ij}^2}\cos\rho_{ij}^2((\kappa_1)^2+(\kappa_2)^2) ds,
 \end{aligned}
 \label{eq:Ptot}
 \end{equation}

where $ds=r^2\sin\theta d\theta d\phi$, $\kappa_1$ and $\kappa_2$ are:
\begin{equation}
\small
\begin{aligned}
 \kappa_1 &= \cos\theta_{ij} \cos\phi_{ij} \mu_1 + \cos\theta_{ij} \sin\phi_{ij} \mu_2  -\sin\theta_{ij} \mu_3\\
\kappa_2 &= -\sin\phi_{ij} \mu_1 +\cos \phi_{ij} \mu_2
\end{aligned}
\end{equation}


\vspace{-0.08in}

 The antenna radiation impedance, $R_{rad}$, based on the measured power is estimated in\cite{impedance} which is adopted here, i.e.,
 \vspace{-0.09in}
\begin{equation}
\small
R_{rad}=\frac{{60mn V_0^2  \eta}}{ \sum\limits_{i=1}^{m} 
    \sum\limits_{j=1}^{n} I_{ij}^2 \pi^2 E_0^2 \chi},
\end{equation}
where $V_0$ denotes the voltage at the edge of input port. By measuring the radiation impedance, the input impedance of the antenna, $Z_{ant}$, will be calculated as:

\begin{equation}
\small
Z_{ant}=R_{rad}\frac{P_{in}}{P_{tot}}+jX_{ant},
\end{equation}
which is simplified to a real impedance under negligible values of $X_{ant}$, a condition that is confirmed in the next section. We consider an array of antennas instead of a single antenna to ensure sufficient bending according to Fig. \ref{fig:package}(d).

\subsection{BER of 6G systems using Bent Antenna Arrays}
To capture the impact of antenna bending on the received power by the receiver, we deploy the Friis formula and the received power $P_r$ will become
\begin{equation}
\small P_r = P_{t_{\text{max}}} \times G_{T,F}G_{r}\times\frac{\lambda}{\left(4\pi d\right)^2}\times\frac{16Z_{PA}^2Z_{ant}^2}{\left(Z_{PA}+Z_{ant}\right)^4},
\label{eq:PR}
\end{equation}
where $P_t$, $G_{T,F}$, $G_r$, $\lambda$, $Z_{PA}$ and $d$ represent the transmitter power, gain of folded transmitter antenna array, unfolded receiver antenna gain, wavelength, power amplifier impedance and the distance between the transmitter and receiver, respectively.
According to the receiver power in (\ref{eq:PR}), for a 6G QAM transceiver, the signal-to-noise ratio on the receiver side is calculated as:
\begin{equation}
   \small SNR_{Rec} = \frac{P_r}{P_{r,n_{in}} + P_{n_{out}}},
\end{equation}
where $P_r$, $P_{r,n_{in}}$, and $P_{n_{out}}$ represent the power of the received signal, the power of the received transmitter noise, and the power of the output noise respectively.  The error-vector magnitude of a QAM constellation is approximated by $1/\sqrt{SNR_{Rec}}$ when number of symbol streams ($T$) is much larger than number of modulation symbols ($N$).
Subsequently, the bit-error-rate will be approximated as a function of modulation order, and EVM by
\begin{equation}
\small
    P_b = \frac{2(1 - \frac{1}{L})}{{\log_2(L)}}  \cdot Q\left(\sqrt{\frac{{3 \log_2(L)}}{{L^2 - 1}} \cdot \frac{2}{{\text{EVM}^2_{RMS} \log_2(M)}}}\right),
\end{equation}
where M and L are mode of modulation and the number of levels in each dimension of the
M-ary modulation system.

\begin{figure}
       \centering
       \includegraphics[trim={0.2cm 0.0cm 0.0cm 0.2cm},clip, width=0.9\columnwidth]{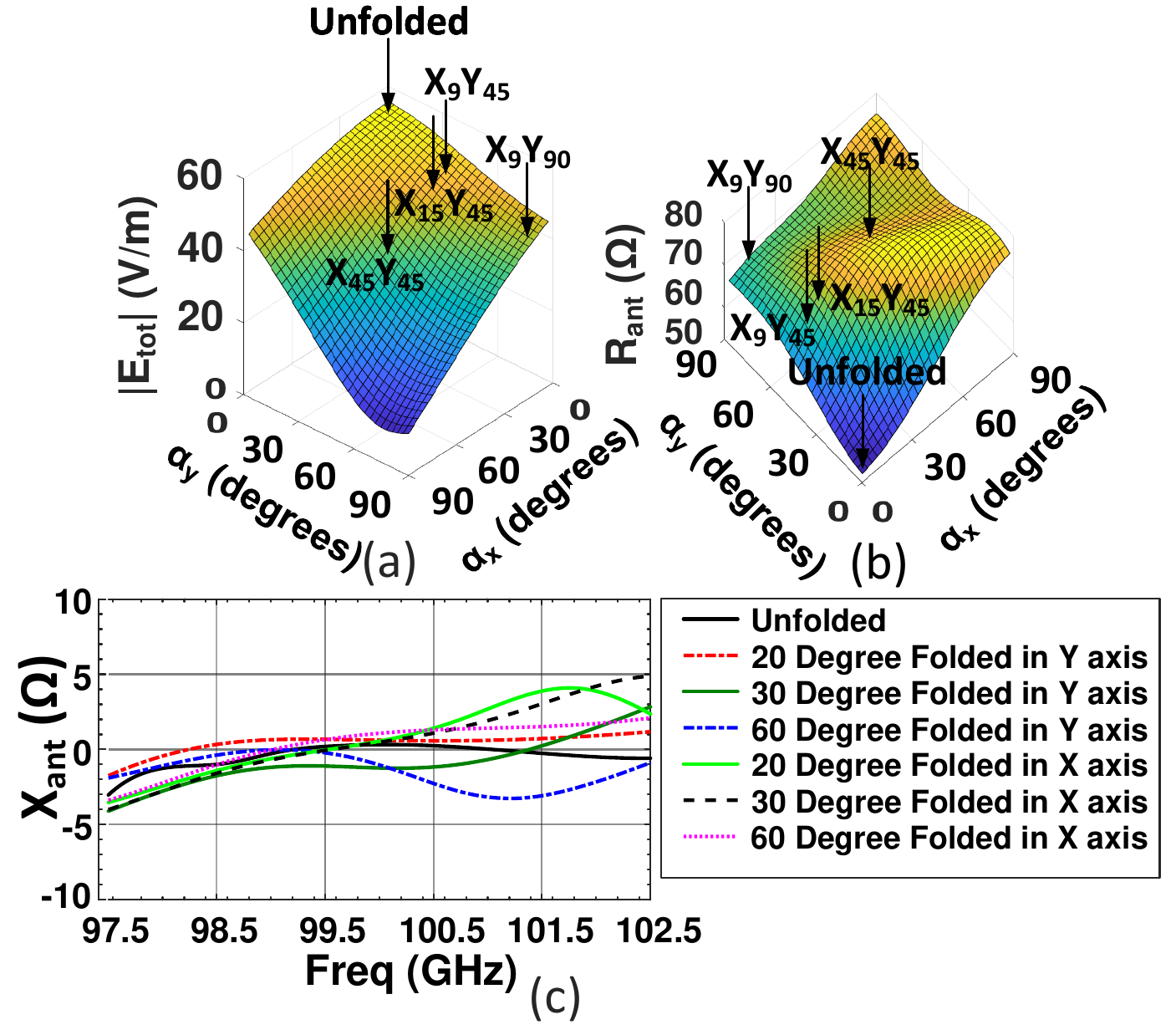}
       \vspace{-0.10in}
       \caption{(a) Analytical values versus simulated results for (a) the magnitude of electric field, and (b) input resistance of the 4$\times$4 array, (c) input reactance variations by folding.}
       \label{fig:EtotZ}
   \end{figure}

       
        

\begin{figure}[t]
    \centering
    \begin{subfigure}[b]{0.40\linewidth} 
        \centering
        \includegraphics[trim={1.8cm 0.0cm 7.5cm 3.5cm},clip, width=\linewidth]{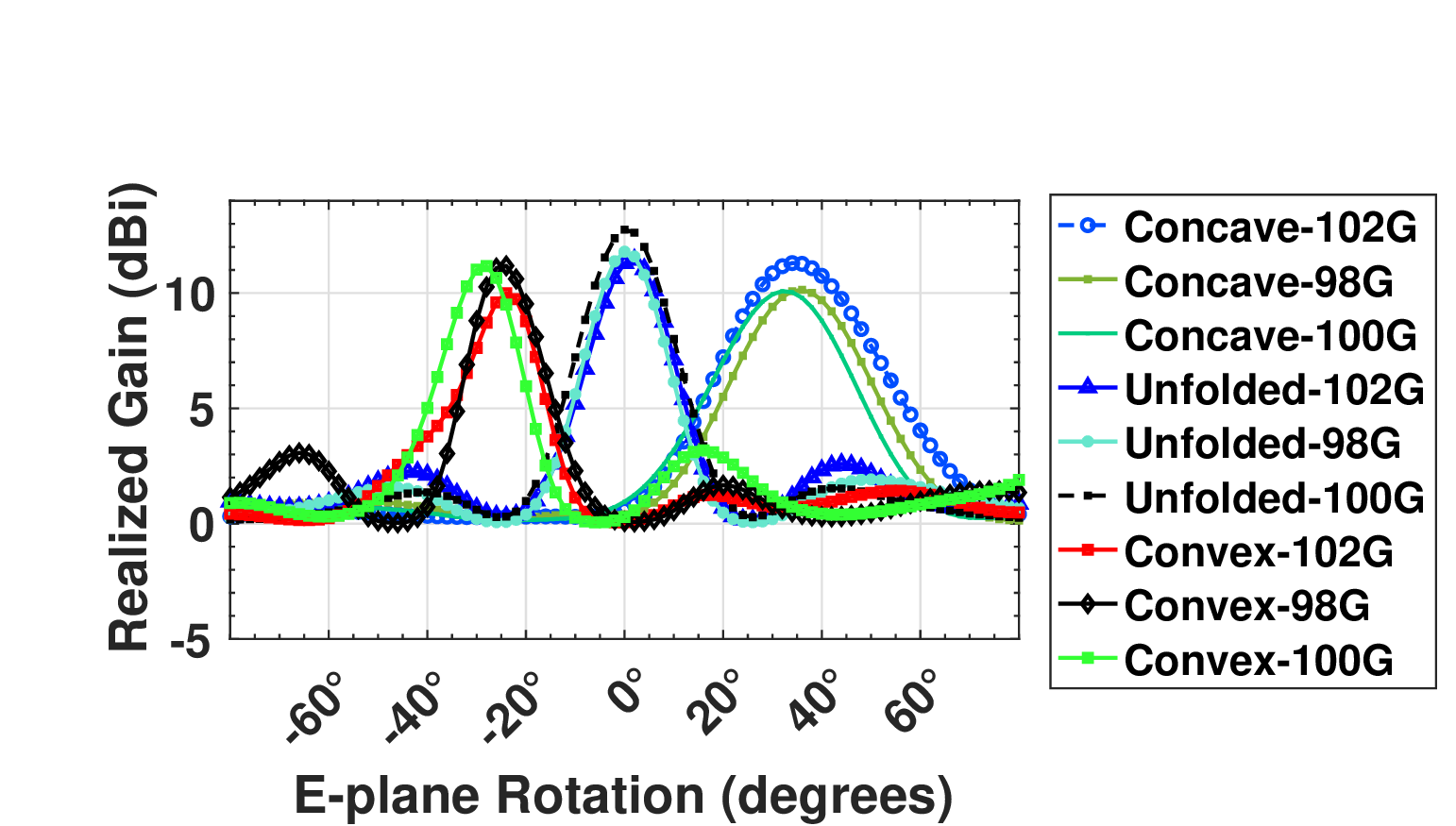}
        \label{fig:subfig1}
    \end{subfigure}
    \hfill
    \begin{subfigure}[b]{0.57\linewidth} 
        \centering
        \includegraphics[trim={0.2cm 0.0cm 2.2cm 3.5cm},clip, width=\linewidth]{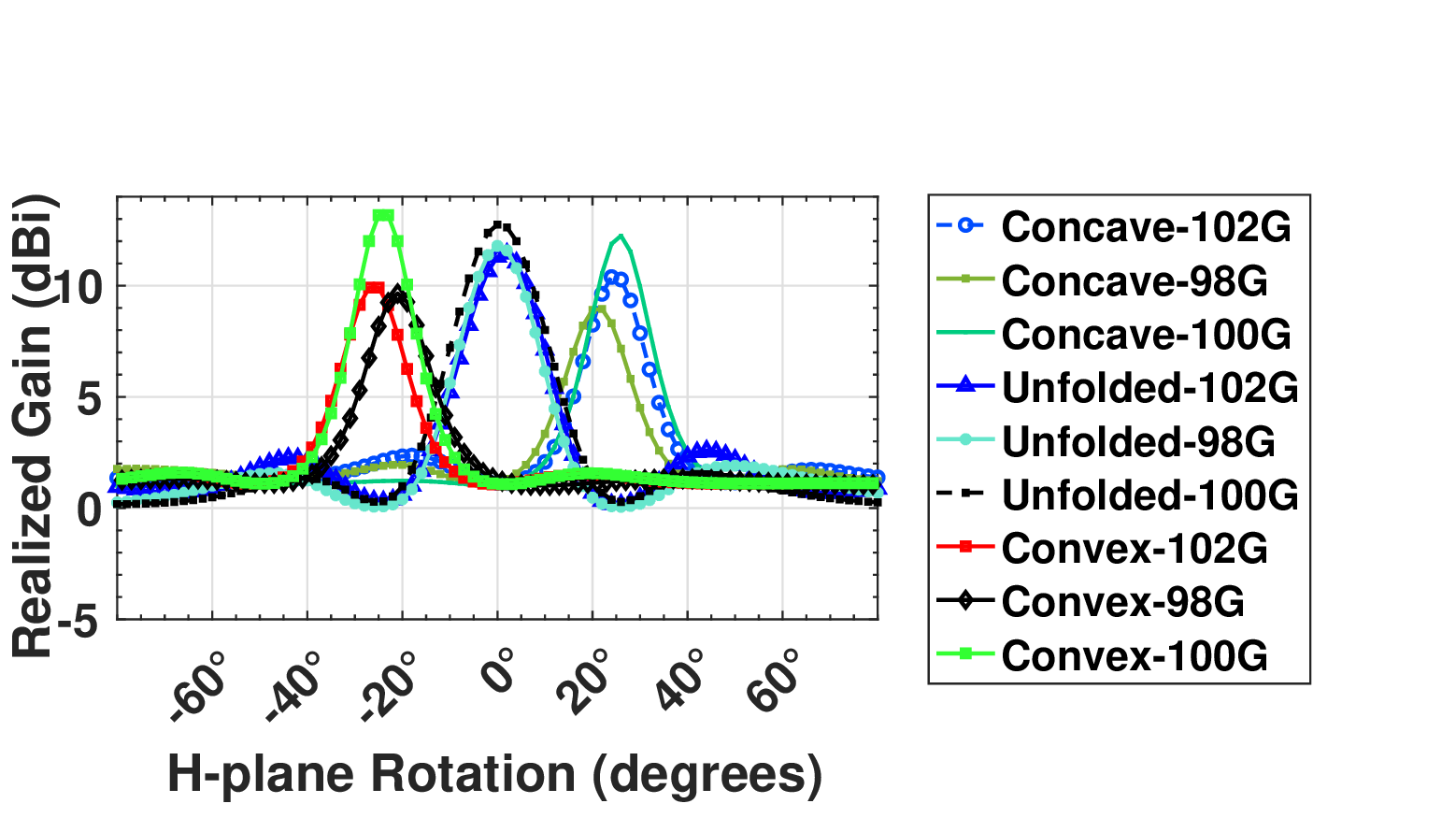}
        \label{fig:subfig2}
    \end{subfigure}
    \vspace{-0.2in}
    \caption{Flexible antenna array beam profile across the bandwidth in (left) E-plane and (right) H-plane.}
    \label{fig:mainfig}
\end{figure}
\setlength{\textfloatsep}{1pt}
 \begin{figure}[t!]
       \centering
       \vspace{-0.9em} 
        \includegraphics[trim={0.2cm 0.0cm 0.2cm 0.0cm},clip, width=0.95\columnwidth]{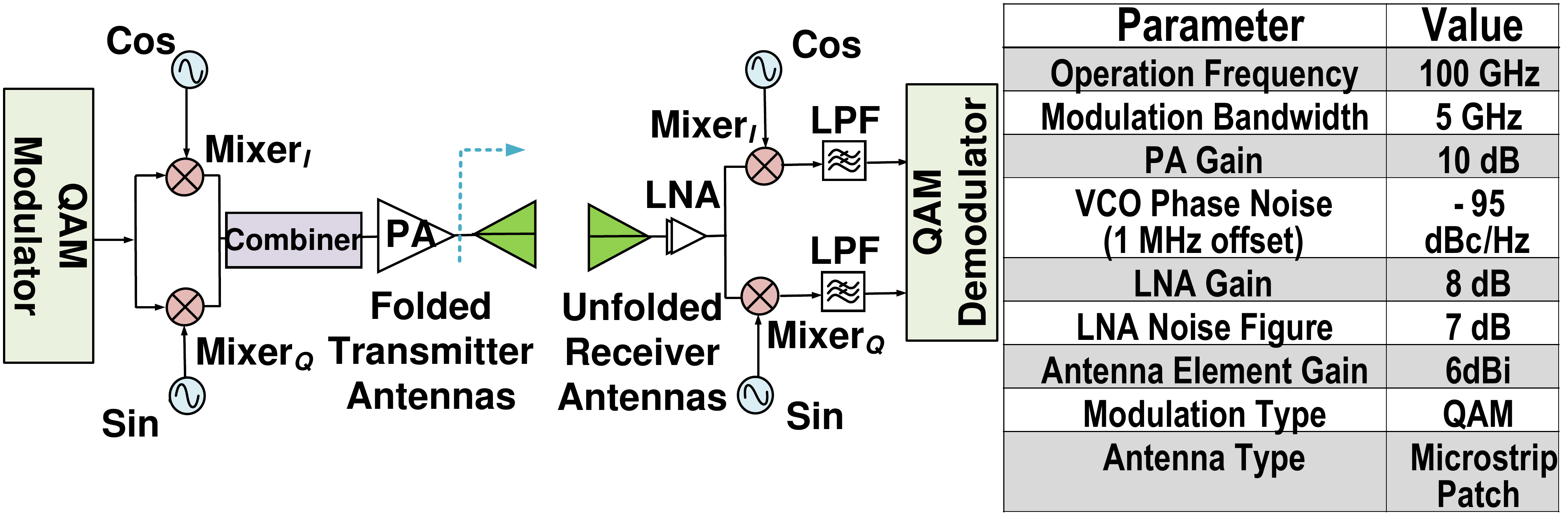}
       \caption{The system level architecture of the simulated 6G wireless system and the design parameters}
       \label{trans}
   \end{figure} 
\begin{figure}[t!]
    \centering
    \vspace{-0.8em} 
    \includegraphics[width=1\linewidth]{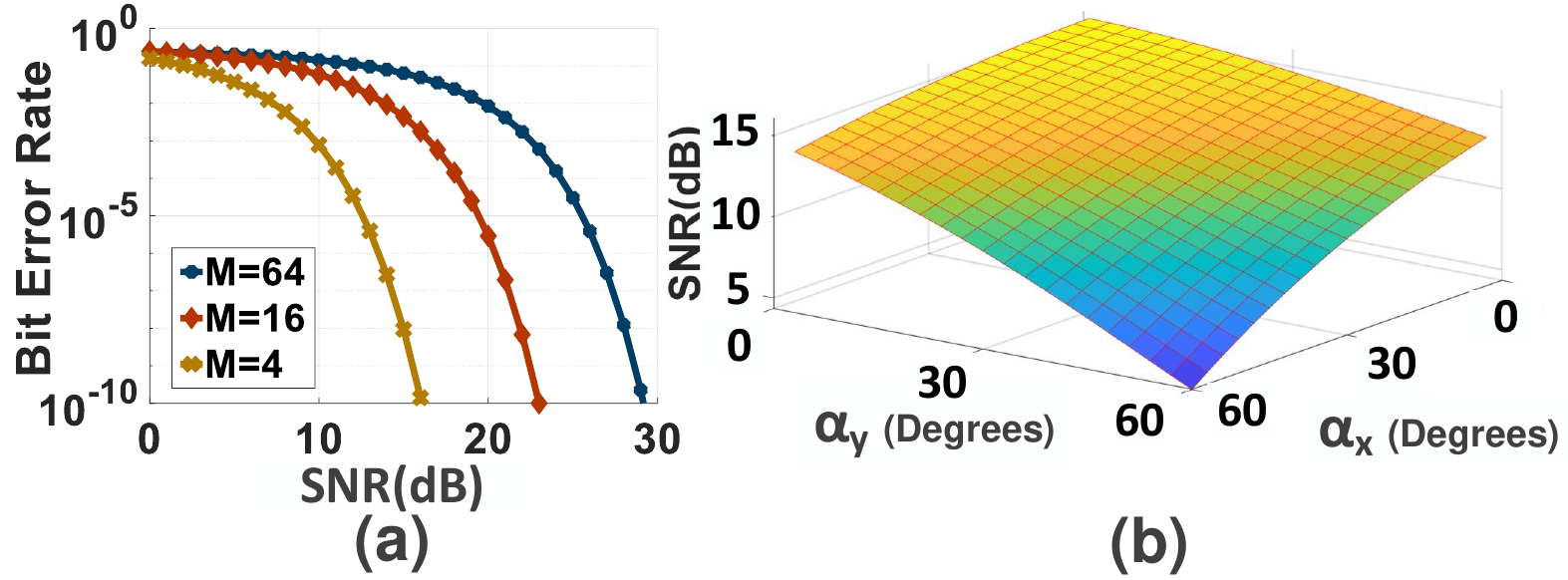}
    \caption{(a) BER based on SNR for different modes of QAM, and (b) SNR variation with various folding degrees along the x- and y- axis.}
    \label{fig:snr}
\end{figure}
\setlength{\textfloatsep}{9pt} 
\vspace{-0.1in}
 \section{Simulation Results}
To assess the analytical derivations in Section II, the $4\times4$ patch antenna designed based on MFLEX modified polyimide material was simulated in Ansys HFSS with four different folding angles along the \textit{x}- and \textit{y}-axis. The angle of bending in the \textit{x}-axis and \textit{y}-axis are denoted as $x_{\xi_1},y_{\xi_2}$ where $\xi_1$ and $\xi_2$ are the angle of bending between the first and last antennas in a row or column along the \textit{x} and \textit{y} directions, respectively.  The simulated folded antenna array versions include $x_9,y_{45}$, $x_9,y_{90}$, $x_{45},y_{45}$, and $x_{15},y_{45}$, in addition to the unfolded case. . As shown in Fig. \ref{fig:EtotZ}(a), the magnitude of radiated electric field changes with the folding angle and for excessively large angles (e.g., $\geq 90^0$), the destructive summation of fields from antennas results in weakened radiation intensity. The input impedance of the antenna array, $R_{ant}$, also changes by the folding degree and increases gradually by the bending angle, Fig. \ref{fig:EtotZ}(b). The total electric field and input resistance for the five scenarios, given as, ($|\vec{E}_{tot}|$,$R_{ant}$), are (51,50), (46,67), (38,62), (30,67), and (46,72), respectively. The simulation results align with the analytical equations. Moreover, $X_{ant}$, the reactive part of input impedance, is simulated for various bending scenarios in the \textit{x} and \textit{y} directions where the variations are negligible and bounded between $\pm$ 5$\Omega$, Fig. \ref{fig:EtotZ}(c). To account for the possible beam squint \cite{squint} associated with the folding, in Fig. \ref{fig:mainfig}, the beam stability across the bandwidth was simulated in HFSS  showing minimal squint in both the E-plane and H-plane, making this design suitable for 6G systems.

The folded transmitter antenna array was simulated within a post-layout CMOS-based 6G wireless transceiver (Fig. \ref{trans}). Digital baseband QAM signals were generated in the DSP, converted to analog via DACs, upconverted to RF using I-Q mixers driven by quadrature LO signals, amplified, and transmitted through the folded array. The unfolded receiver antenna amplified, downconverted, and filtered the signal, with a digital QAM demodulator extracting the bits. Fig. \ref{fig:snr}(a) shows that higher QAM spectral efficiency requires increased SNR to maintain BER \cite{EVM}. Excessive folding reduces SNR, but both simulated and analytical results (Table I) confirm the folded array's viability for 4-QAM and 16-QAM modulations. For 64-QAM, the transmitter achieves a BER within $O(10^{-2})$ for small folding angles. 
The slight discrepancy between the simulation results and the analytical equations in Table I arises from simplified electric field assumptions, the omission of imaginary impedance, and the approximations used for the EVM of QAM symbols.

\begin{table}[t!]
    \centering
      \caption{Simulated and analytical BER for different QAM modes}
    \includegraphics[trim={0cm 0cm 0cm 0cm},clip, width=0.95\columnwidth]{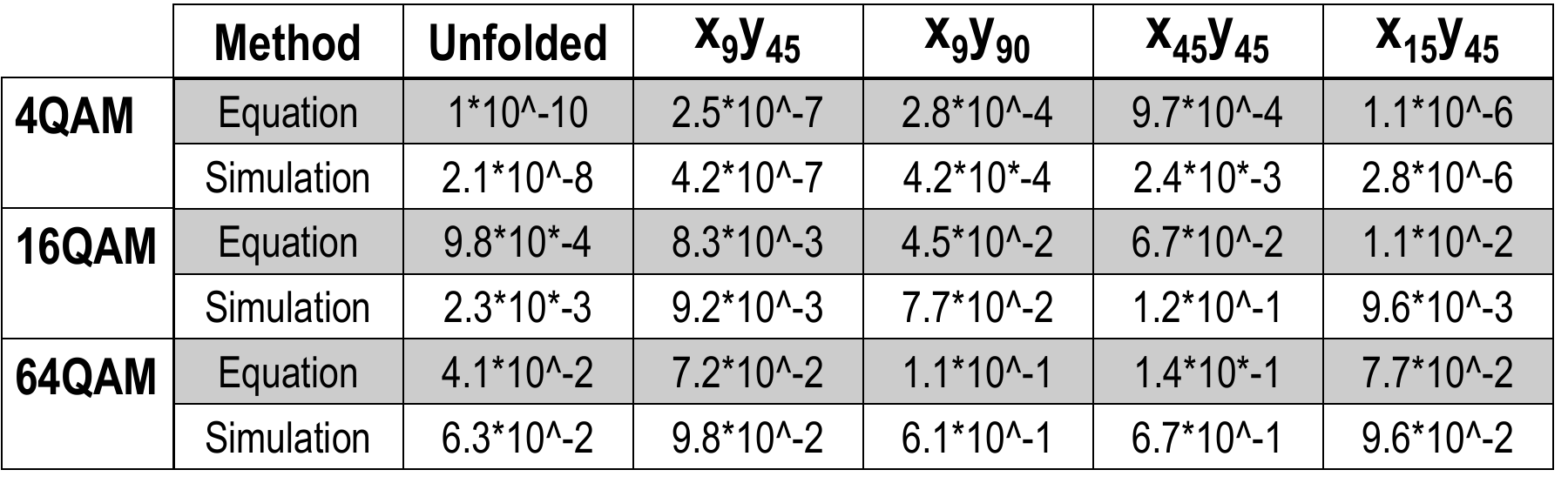}
  
    \label{fig:enter-label}
\end{table}

\begin{table}[ht]
\centering
\caption{\footnotesize  of Beam Rotation Techniques for 6G}
\fontsize{6.7}{6.7}\selectfont
\setlength{\tabcolsep}{1pt} 
\renewcommand{\arraystretch}{1.5} 
\resizebox{\columnwidth}{!}{ 
\begin{threeparttable}
\begin{tabular}{|>{\centering\arraybackslash}p{1cm}|>{\centering\arraybackslash}p{1cm}|>{\centering\arraybackslash}p{1cm}|>{\centering\arraybackslash}p{1cm}|>{\centering\arraybackslash}p{1cm}|>{\centering\arraybackslash}p{1cm}|>{\centering\arraybackslash}p{1.2cm}|}
\hline
\textbf{Reference} & \cite{18} & \cite{27} & \cite{25} & \cite{phase_shifter_Wooram} & \cite{30} & \textbf{This work} \\ \hline
\textbf{Frequency} & 85 GHz & 86 GHz & 64 GHz & 145 GHz & 160 GHz & \textbf{100 GHz} \\ \hline
\textbf{Type} & Active PS\tnote{*} & VM\tnote{**} & Passive PS & Passive PS & Active PS & $\#$Flex Ant \\ \hline
\textbf{IL (dB)} & 7.2 dB & 4.99 dB & 16.3 dB & 11.5 dB & 4.5 dB & \textbf{2 dB} \\ \hline
\textbf{$P_{dc}$(mW)} & 150*** & 0 & 0 & 0 & 50 & \textbf{0} \\ \hline
\end{tabular}
\begin{tablenotes}
\item[*] PS = phase shifter, **VM = vector modulator,  *** = with PA,  $\#$Flex Ant= Flexible Antenna,  IL=Insertion Loss, $P_{dc}$ is the DC power consumption
\end{tablenotes}
\end{threeparttable}
}

\label{tab:antenna-comparison}
\end{table}

\vspace{-0.2in}
\section{Conclusion}
This work examined the 2-D curvature effects of flexible antenna arrays on the radiation properties as well as communication performance in a 6G wireless system. In addition a beam rotation without peak gain degradation is confirmed up to 60 degrees, without a noticeable beam squint. Moreover, the deployment of the folded antenna array in a CMOS-based QAM transceiver is studied to confirm low BER across the desired bending degrees of the antenna array. As compared against phase shifting techniques in TABLE II, by designing MIMO phased arrays deploying these flexible antennas, the beam alignment can be extended in 2 dimensions beyond the reach of phase shifters while reducing the insertion loss associated with conductor and dielectric losses and slight impedance mismatches due to folding.



 
%

\bibliographystyle{IEEEbib}
\bibliography{reference2.bib}
\end{document}